\documentclass[12pt]{article}
\pdfoutput=1
\usepackage{color}
\usepackage{amsmath,amssymb}
\usepackage{graphicx}
\usepackage{hyperref}
\usepackage{titlesec}
\usepackage{amsmath}
\usepackage{ae}
\usepackage{units}
\usepackage{icomma}
\usepackage{color}
\usepackage[usenames,dvipsnames,svgnames,table]{xcolor}
\usepackage{graphicx}
\usepackage{amssymb}
\usepackage[left=2.55cm,right=2.55cm , top=4cm, bottom=4cm]{geometry}
\makeatletter

\renewcommand\section{\@startsection{section}{1}{\z@}
                                   {-3.5ex \@plus -1ex \@minus -.2ex}
                                   {2.3ex \@plus .2ex}
                                   {\normalfont\large\bfseries}}
\renewcommand\subsection{\@startsection{subsection}{2}{\z@}
                                   {-3.25ex\@plus -1ex \@minus -.2ex}
                                   {1.5ex \@plus .2ex}
                                   {\normalfont\normalsize\bfseries}}
\renewcommand\subsubsection{\@startsection{subsubsection}{3}{\z@}
                                   {-3.25ex\@plus -1ex \@minus -.2ex}
                                   {1.5ex \@plus .2ex}
                                   {\normalfont\normalsize\bfseries}}
\renewcommand\paragraph{\@startsection{paragraph}{4}{\z@}
                                   {3.25ex \@plus1ex \@minus.2ex}
                                   {-1em}
                                   {\normalfont\normalsize\bfseries}}

\makeatother

\newcommand{\bc}{\begin{center}}
\newcommand{\ec}{\end{center}}
\newcommand{\beq}{\begin{equation}}
\newcommand{\eeq}{\end{equation}}
\newcommand{\bea}{\begin{align}}
\newcommand{\eea}{\end{align}}

\newcommand{\Gf}{\Gamma_4}
\newcommand{\G}{\Gamma}

\newcommand{\Spin}{\rm Spin}

\newcommand{\id}{\hbox{1\kern-.27em l}}

\newcommand{\ad}{{\rm ad}}

\begin{document}

\pagestyle{empty}

\begin{center}

\vspace*{30mm}
{\LARGE Five-dimensional topologically twisted maximally supersymmetric Yang-Mills theory}

\vspace*{30mm}
{\large Louise Anderson }

\vspace*{5mm}
Department of Fundamental Physics\\
Chalmers University of Technology\\
S-412 96 G\"oteborg, Sweden\\[3mm]
{\tt louise.anderson@chalmers.se}

\vspace*{30mm}{\bf Abstract:}\\
\end{center}
Herein, we consider a topologically twisted version of maximally supersymmetric Yang-Mills theory in five dimensions which was introduced by Witten in 2011. We consider this theory on a five manifold of the form $M_4\times I$ for $M_4$ an oriented Riemannian four manifold. The complete and unique action of the theory in bulk is written down and is shown to be invariant under two scalar supersymmetries.

\newpage \pagestyle{plain}

\section{Introduction}

In four dimensions, there are many examples of topological field theories.
 $\mathcal{N}=2$ super-Yang-Mills theory in four dimensions admits one unique topological twisting \cite{Witten1988}, whereas $\mathcal{N}=4$ super-Yang-Mills admits three inequivalent twistings \cite{Yamron, VafaWitten, KapustinWitten, Marcus1995}. 
 In five dimensions however, the situation is slightly different and not as well-studied.  Maximally supersymmetric Yang-Mills theory considered on a general five-manifold of Euclidean signature, $M_5$, has both R-symmetry- and Holonomy group (of $M_5$) equal to $SO(5)$. Hence there exist a unique topological twisting of this theory that will give one scalar and nilpotent supersymmetry charge. However, if one considers a five-manifold not so general but rather on the form
 \beq
 \label{mnfld}
 M_5=M_4 \times I
, \eeq
with $I$ some one-dimensional manifold and $M_4$ a Riemannian four-manifold of Euclidean signature, the holonomy group is reduced to $SO(4)$, and the theory may admit several topological twistings. 

One of these result in a topological field theory which is invariant under two scalar nilpotent supersymmetries, and is the theory that will be considered herein. This twisting of five-dimensional maximally supersymmetric Yang-Mills theory can be described as the five-dimensional analog of the four-dimensional geometric Langlands-twist that was first mentioned in \cite{Yamron}, and then shown to have applications to the geometric Langlands program in \cite{KapustinWitten}. This five dimensional topological field theory was first considered in \cite{Witten2011} with the original motivation being the interesting applications to Khovanov homology for knots \cite{Khovanov} it was shown to possess, something that was further studied in \cite{Witten2011-2}.  
In order to make contact with knot theory, and more particularly Khovanov homology, the four-manifold $M_4$ must be a product of a three manifold in which the knots are embedded and another one-dimensional manifold with a boundary. 
 
  This relatively new topological field theory have been the interest of several papers since its introduction \cite{Witten2011-2, AndersonHenningson}, and its four-dimensional analog has been even further studied (amongst others in \cite{ Henningson2011, GaiottoWitten2011, Henningson2011-2,Mikhaylov2012}, and from a lattice supersymmetry perspective in for example \cite{Marcus1995,Kaplan2005,Unsal2006}). However, focus in these works has mainly been on the bosonic aspects of the theory, or probably most heavily on the localisation equations since the connection to knot theory and Khovanov homology lies therein. In this setting, the knots are encoded in subtle boundary conditions for these elliptic differential equations on $M_5$ \cite{Witten2011}.  Much progress have been made in the field,  but as previously mentioned only a part of the theory has been studied in detail.  There are large areas that still remain unexplored.

In this paper, we shall focus on other aspects of this theory than the applications to knot theory, thus the extra requirements upon the manifold that were imposed in order to make contact with this will here be unnecessary. We shall instead consider the theory on a more general five-manifold on the form given in equation \ref{mnfld}. Since the twist was constructed to give a theory containing two scalar nilpotent supersymmetries, it is obvious that it indeed produces a topological field theory when $M_4$ has vanishing curvature. However, it is not as straight forward to see that this is true even in the case when $M_4$ is curved. It is thus important in order to understand this theory better to find the explicit expression of the action, even in the case when $M_4$ is curved. This has not yet been done. However, it should be noted that some terms in this action were written down in \cite{Witten2011} during a discussion regarding the equations of motion of the theory.

 In this paper, we find the expression for the complete action when $M_4$ is a  general Riemannian four-manifold of Euclidean signature, and this is concluded to be unique. 
 It is furthermore shown to be invariant under both scalar, nilpotent supersymmetries in bulk. 
 This is done by first describing the twist in greater detail as well as the field content of the theory. After this, the action in flat space is computed and shown to be invariant under the two scalar supersymmetries in bulk. Finally, the case when $M_4$ is curved is considered.

  Throughout the paper, we shall only occupy ourselves with the theory in bulk, details of the boundary behaviour can for example be found in \cite{Witten2011} and \cite{GaiottoWitten2011}.  We shall make some brief comments about the importance of the boundary at the end of the paper but not dwell on it further at the moment.

Finding an expression for the action is an important step in order to understand this topological field theory, but many questions still remain. It may for example be interesting to investigate if the action presented herein is $Q$-exact. My belief is that this is not the case since the four-dimensional analog of this theory does not have a $Q$-exact action \cite{KapustinWitten}, however, it would be interesting to study this in greater detail. It may also be interesting to investigate the surface terms in the action that have not been presented herein, and study their behaviour and properties.

\section{The Twist}

The five-dimensional maximally supersymmetric Yang-Mills theory discussed above is considered  as a dimensional reduction from ten dimensions and we work in overall Minkowski signature. In a ten-dimensional notation, the theory may famously \cite{Brink1976} be described by the action 
\begin{align}
\label{ten_dim_action}
\mathcal{S} = 
\int _{M_4 \times I} \text{d}^4x \text{d}y \text{Tr } \left(
-\frac{1}{4}F_{MN}F^{MN} +\frac{i}{2}\bar{\lambda} D_M \G^M \lambda
\right)
.\end{align}
This action is invariant under the supersymmetry transformations below:
\begin{align}
\label{10_dim_bos}
\delta A_I = i \bar{\varepsilon} \G_I \lambda
\end{align}
\begin{align}
\label{ferm_variations}
\delta \lambda = & \frac{1}{2}F_{IJ}\G^{IJ}\varepsilon
.\end{align}
This will be our starting point to obtain the action of the twisted theory.

 There have previously been comments that topological twisting only works in Euclidean signature, but in this specific case, this is not entirely true. Minkowski signature is handleable under the condition that the time-like direction is chosen to lay in the direction along the interval $I$. In this manner we restrict ourselves to a compact subgroup of the non-compact Lorentz group of $M_5$, namely the subgroup consisting of rotations on $M_4$. This subgroup will be compact under the only requirement that $M_4$ is an oriented Riemannian manifold of Euclidean signature. Thus the overall Minkowski signature is no hinderance to perform the topological twisting, which corresponds to a homomorphism from the $\Spin (4)$ holonomy group of $M_4$ to the $\Spin (5)$ $R$-symmetry group of the Yang-Mills theory under which the spinor representation ${\bf 4}$ of $\Spin (5)$ decomposes as a direct sum ${\bf 2} + {\bf 2}$ of two chiral spinor representations of $\Spin (4)$.  This twisting is described in greater detail in the table below.

\bc
\begin{tabular}{ c c @{ \ } c @{}c @{ }c}
              & $SU(2)_l\times SU(2)_r \times U(1) \times SU(2)_R$  & $\xrightarrow[]{twist}$ & $ SU(2)_l\times
              SU(2)' \times U(1)$ & \\
  $A_\mu$     & $\mathbf{(2,2,1)^0}$              &    &
   $\mathbf{(2,2)^0} $  & $A_\mu$
   \\
  $A_y$       & $\mathbf{(1,1,1)^0}$              &      &
   $\mathbf{(1,1)^0}$  & $A_y$
    \\
  $\Phi_I$    & $\mathbf{(1,1,1)^{+1}}\oplus \mathbf{(1,1,1)^{-1}} \oplus \mathbf{(1,1,3)^{0}}$     &      &
   $\mathbf{(1,1)^{+1}}\oplus \mathbf{(1,1)^{-1}} \oplus \mathbf{(1,3)^{0}} $ 
   & $\sigma, \bar{\sigma}, B_{\mu \nu}$
   \\
  $\lambda_\alpha$ & $\mathbf{(1,2,2)^{+1/2}}\oplus \mathbf{(1,2,2)^{-1/2}} $  & &  $\mathbf{(1,1)^{\pm
  1/2}}\oplus
  \mathbf{(1,3)^{\pm 1/2}} $
  & $\eta, \tilde{\eta}, \chi_{\mu \nu}, \tilde{\chi}_{\mu \nu}$
   \\
             & $\oplus \mathbf{(2,1,2)^{+1/2}}\oplus \mathbf{(2,1,2)^{-1/2}} $ & &  $\oplus \mathbf{(2,2)^{\pm
             1/2}}$
             & $\psi_\mu, \tilde{\psi}_\mu$
              \\
\end{tabular}
\ec
The left hand side here contains the fields and the respective representations in which they live after dimensional reduction from ten dimensions, whereas the right hand side contains the same information for the twisted theory. $I$ takes the values ${5,6,7,8,9}$, so  $\Phi_I$  thus denotes the ten-dimensional gauge field in the $5,6,7,8,9$-directions. The dimensions of the representations of the $SU(2)$s are represented by bold face numbers, and the charge under the U(1) is denoted by a superscript. The twisting here replaces $SU(2)_r \times SU(2)_R$ with $SU(2)'$, which is the diagonal group of $SU(2)_r \times SU(2)_R$ .

 After such a twisting, the bosonic degrees of freedom can be described by fields on the four manifold that in addition depends on the linear coordinate $y$ along $I$. These will be the gauge connection $A_\mu$ with field strength $F_{\mu \nu}$, a self-dual  (with respect to the orientation and Riemannian structure of $M_4$)  two-form $B_{\mu \nu}$ and a complex scalar $\sigma$. These will all take their values in the vector bundle $\ad(E)$ associated to the  gauge bundle $E$ via the adjoint representation of the gauge group $G$.
Furthermore, the fermionic degrees of freedom will after the twisting be given by two zero-forms ($\eta, \tilde{\eta}$), two one-forms ($\phi_\mu, \tilde{\phi}_\mu $), and two self-dual two-forms ($\chi_{\mu \nu}, \tilde{\chi}_{\mu \nu}$), which will also take values in $\ad(E)$. The two fermionic forms of a certain degree will be distinguished by opposite charge under the $U(1)$  of the $\Spin(5)$ $R$-symmetry group of the Yang-Mills theory that is left untouched by the twisting. These fields are all summarised in the table below.

\bc
\begin{tabular}{ c c c c c c}
     Bosonic Fields           &       &                        & Fermionic Fields                        &       &
     \\
     $A_\mu$                  & $\in$ & $\Omega^1(M_4, \ad(E))$    & $\eta, \tilde{\eta}$                    & $\in$ &
     $\Omega^0(M_4, \ad(E))$\\
     $A_y$                    & $\in$ & $\Omega^0(M_4, \ad(E))$    & $\psi_\mu, \tilde{\psi_\mu}$              & $\in$
     & $\Omega^1(M_4, \ad(E))$\\
     $\sigma , \bar{\sigma}$  & $\in$ & $\Omega^0(M_4, \ad(E))$    & $\chi_{\mu\nu}, \tilde{\chi_{\mu\nu}}$  & $\in$ &
     $\Omega^{2+}(M_4, \ad(E))$\\
     $B_{\mu\nu}$             & $\in$ & $\Omega^{2+}(M_4, \ad(E))$ &                                         &       &
     \\
\end{tabular}
\ec 

Before we proceed any further, we shall spend some time on clarifying some details in our notation. 
The bosonic self-dual two form in the twisted theory is simply obtained by reinterpreting the original ten-dimensional gauge fields in the $5,6,7$-direction as components of a self-dual two form according to the relations below:
\begin{align}
\label{Definition_B}
B_{0 i} &=\phi_i
\\ \nonumber 
B_{ij}&=\epsilon_{ijk}\phi^i
.\end{align}
Here, $i,j \in \{1,2,3\}$, and we define $\phi_i=\Phi_{i+4}$.
The statement that $B_{\mu \nu}$ is self-dual is as usual equivalent to saying that $B_{\mu \nu}$ satisfies $B^{\mu \nu}\epsilon_{\mu \nu \kappa \lambda}  = 2B_{\kappa \lambda}$. Furthermore, the self-dual part, denoted $\Omega^+_{\mu \nu}$, of a general two-form $\Omega_{\mu \nu}$ can be written as $\Omega^+_{\mu \nu} = \frac{1}{2}(\Omega_{\mu \nu} +\frac{1}{2} \epsilon_{\mu \nu \rho \sigma} \Omega^{\rho \sigma})$. This notation will be helpful in the coming calculations.

Furthermore, as also done in \cite{Witten2011}, it will be convenient to define a product on the space of self-dual two-forms according to
\begin{align}
\label{Definition_B_times_B}
(B \times B)_{\mu \nu} =\sum_{\tau=0}^3 [B_{\mu \tau}, B_{\nu \tau}]
.\end{align}
It is quite straight-forward to check that if $B \in \Omega^{2+}(M_4, \ad(E))$, then $B\times B \in \Omega^{2+}(M_4, \ad(E))$.

Furthermore, the complex valued zero-forms $\sigma$ (with complex conjugate $\bar{\sigma}$) is obtained by  reinterpreting the gauge fields in the $8,9$-direction, so we find:
\begin{align}
\label{Definition_Sigmas}
\sigma &=\frac{\phi_8 -i \phi_9}{\sqrt{2}}
\\ \nonumber 
\bar{\sigma} &=\frac{\phi_8 +i \phi_9}{\sqrt{2}}
.\end{align}

\subsection{Twisting of the supersymmetries and fermionic fields}
The supersymmetries transform in the spinor representation of the gauge group before the twist, as is the case for the fermions. Thus after the twist,  as for the fermions,  two of these will be scalar under the two $SU(2)$'s remaining after the twist ($SU(2)_l$ and $SU(2)'$), with opposite $U(1)$-charge. These two supersymmetries will live in the representations  $(\mathbf{1}, \mathbf{1})^{+1/2}$ and  $(\mathbf{1}, \mathbf{1})^{-1/2}$ respectivly.
 Let the $(\mathbf{1}, \mathbf{1})^{+1/2}$ be generated by the constant spinor $e^{+}$. Then $(\mathbf{1}, \mathbf{1})^{-1/2}$ is generated by $e^{-}=\frac{1}{\sqrt{2}}\Gf \Gamma_{8-i9} e^{+}$, where $\G_{8- i9}=\frac{1}{\sqrt{2}}(\G_8 - i\G_9)$.
Since the twisted theory is invariant under any linear combination of the two supersymmetries, this can be written as it being invariant under a supersymmetry with parameter
\beq
\label{susys}
\varepsilon = u e^+ + v e^- = (u+v \frac{1}{\sqrt{2}}\Gf \Gamma_{8-i9})e^+
,\eeq
where $u$ and $v$ are grasmannian odd with $U(1)$-charges $\pm \frac{1}{2}$ respectively. Thus we have here an entire family of supersymmetries under which the theory is invariant, described by the relations between $u$ and $v$. For completeness, one can here notice that the boundary conditions of the theory is half-BPS, thus fixing the relationship between $u$ and $v$ such that $u = \pm v$ \  \cite{Witten2011}. 
However, as stated previously we will here consider the theory in bulk and must thus still consider general $u$ and $v$. 

In the same way as previously with $(\mathbf{1}, \mathbf{1})^{\pm 1/2}$ , we can also find base elements spanning  $(\mathbf{2}, \mathbf{2})^{\pm 1/2}$ and  $(\mathbf{1}, \mathbf{3})^{\pm 1/2}$ respectively.   These will be generated by  $e^{\pm}_\mu = \Gf \Gamma_\mu e^{\pm}$ and
 $e^{\pm}_{\mu \nu} = \Gamma_{\mu \nu} ^+ e^{\pm}$ respectively. (Recall that with our notation, $ \Gamma_{\mu \nu} ^+$ denotes the self-dual part of $ \Gamma_{\mu \nu} $).This allows us to write the
ten-dimensional fermionic fields in terms of the twisted five-dimensional fields as follows:
\begin{align}
\label{lambda}
\lambda=
\left( \eta+ \frac{1}{\sqrt{2}}\Gf \Gamma_{8-i9}\tilde{\eta} + 
\G_\kappa \Gf \psi^\kappa- \frac{1}{\sqrt{2}}\G_\kappa \Gamma_{8-i9}\tilde{\psi}^\kappa 
-\frac{1}{4} \G_\kappa \G_\lambda \chi^{\kappa \lambda}
-\frac{1}{4}\frac{1}{\sqrt{2}}\G_\kappa \G_\lambda\Gf \Gamma_{8-i9}\tilde{\chi}^{\kappa \lambda} \right) e^{+}
.\end{align}
The self-duality property is here placed upon the coefficients $\chi^{\kappa \lambda}$, thus eliminating the need to use the self-dual part of  $\Gamma_{\mu \nu} $ only. This will facilitate coming calculations.

In order to obtain the fermionic part of the action for the theory, one must also write down the Dirac conjugate of $\lambda$;  $\overline{\lambda}$. This can easily be done by using that for a general spinor $x$, its Dirac conjugate is given by $\bar{x}=x^* \Gf$, where the star denotes complex conjugation. Thus we will have:
\begin{align}
\label{lambda_bar}
\overline{\lambda}=
\bar{e}^{+} \left( \tilde{\eta}- \frac{1}{\sqrt{2}}\Gf \Gamma_{8+i9}\eta- \G_\kappa \Gf \tilde{\psi}^\kappa+ \frac{1}{\sqrt{2}}\G_\kappa \Gamma_{8+i9}\psi^\kappa 
-\frac{1}{4}  \G_\lambda \G_\kappa \tilde{\chi}^{\kappa \lambda}
-\frac{1}{4}\frac{1}{\sqrt{2}}\Gamma_{8+i9}\Gf  \G_\lambda\G_\kappa\chi^{\kappa \lambda} \right)
.\end{align}
In this calculation, the Minkowski signature will have an effect, and we must note that with our conventions we have $\Gf \Gf=-1$.

\subsection{Gamma matrix gymnastics}
The conditions that our supersymmetries in equation \ref{susys} are scalar under both of the remaining $SU(2)$'s after the twist can be written down explicitly by considering the generators of these groups. We denote the generators of the twisted, diagonal subgroup $SU(2)'$ with $\sigma_D$ and the generators of the original $SU(2)_l$ with $\sigma_l$, as:
\begin{align}
\sigma_l^x =& \frac{i}{2}(\G_0 \G_1 - \G_2 \G_3) \\ \nonumber
\sigma_l^y =& \frac{i}{2}(\G_0 \G_2 - \G_3 \G_1) \\ \nonumber
\sigma_l^z =& \frac{i}{2}(\G_0 \G_3 - \G_1 \G_2) \\ \nonumber
\sigma_D^x =& -\frac{i}{2}(\G_0 \G_1 + \G_2 \G_3+ 2\G_5 \G_6) \\ \nonumber
\sigma_D^y =& -\frac{i}{2}(\G_0 \G_2 + \G_3 \G_1+ 2\G_6 \G_7) \\ \nonumber
\sigma_D^z =& -\frac{i}{2}(\G_0 \G_3 + \G_1 \G_2+ 2\G_7 \G_5) 
.\end{align}
The condition that $e^+$ is invariant under both of these subgroups is equivalent to $e^+$ being anhilated by all of the above generators.   This condition can be written on a slightly easier form, namely:
\begin{align}
\label{su2_inv_e}
(\G_i \G_j+\G_{i+4}\G_{j+4}) e^+ &= 0
\\ \nonumber
(\G_0 \G_i -\frac{1}{2} \epsilon_{i j k} \G^j \G^k) e^+ &= 0
,\end{align}
where $i\in \{1,2,3\}$.

In addition to requiring that our constant base element $e^+$ is invariant under the two $SU(2)$'s that exist after the twisting (equation \ref{su2_inv_e}), one can also find other relations for how the gamma matrices act on the spinor $e^+$. Firstly, $e^+$ will be anti-chiral in four dimensions.  It will also be chiral in a ten-dimensional sense, such that $\G_{11}=\G_0...\G_9$ will leave $e^+$ invariant.  Furthermore, we know of the properties it will have under transformation of the $U(1)$, and from all this, one can deduce how the gamma matrices in the $5,6,7$-directions will act on it. All of these properties can be summarised as:
\begin{align}
\label{basic_prop_gammas}
\G_0 \G_1 \G_2 \G_3 e^+=& - e^+ \\ \nonumber
\G_0 \G_1 \G_2 \G_3 \G_4 \G_5 \G_6 \G_7 \G_8 \G_9 e^+=&   e^+  \\ \nonumber
\G_8 \G_9 e^+ =&  i e^+  \\ \nonumber
 \G_{i+4} e^+  =& -\frac{i}{2}\epsilon_{i j k} \G_j\G_k\Gf e^+ 
.\end{align}
There is however some freedom left that we have not used yet. That is the normalisation of $e^+$.  This will lack any physical meaning and result in only an overall scaling of the final action, so in this work we will make the simplest choice such that
\begin{align}
\bar{e^+}\Gf e^+ = &1
.\end{align}
We can now deduce what will happen to expressions on the form $\bar{e}^+ \G_{...} e^+$, which will be needed in order to compute the action of the theory. Any odd number of gamma matrices between $\bar{e}^+$ and $e^+$ will trivially be zero by chirality. For any even number of gamma matrices, one can from the above relations compute the quantity. The most important cases that shall be needed are:
\begin{align}
\label{gammas_between_e}
\bar{e^+}\Gf  \G_\mu \G_\nu e^+ =& \delta_{\mu \nu} \\ \nonumber
\bar{e^+}\Gf  \G_\mu \G_\nu \G_\kappa \G_\lambda  e^+ =&-\epsilon_{\mu \nu \kappa \lambda} + \delta_{\mu \nu} \delta_{\kappa \lambda} + \delta_{\nu\kappa}\delta_{\mu\lambda}-\delta_{\mu\kappa}\delta_{\nu\lambda}
.\end{align}

\subsection{Some useful formulas}

It will in the coming calculations be convenient to note some useful relations before we start. One such is the following:
\begin{align}
\label{twoformprop}
D_{(\nu}D_{\gamma)}  B_{\mu \gamma} \chi^{\mu \nu} =&  \frac{1}{4} D_{\gamma}D^\gamma B_{\mu \nu} \chi_{\mu \nu}
.\end{align}
This can be seen if we study the possible ways of creating a scalar from two self-dual two-forms and the symmetric part of the product between two vectors. The representations of the tensor product between two self-dual two forms  in $SU(2) \times SU(2)$ will be given by
\beq
(\mathbf{1}, \mathbf{3}) \times  (\mathbf{1}, \mathbf{3}) =\left[ (\mathbf{1}, \mathbf{1}) \oplus (\mathbf{1}, \mathbf{5}) \right]_{sym} \oplus \left[ (\mathbf{1}, \mathbf{3}) \right]_{antisym}
.\eeq
Similarly, the symmetric part of the tensor product between two vectors will be obtained as:
\beq
\left[ (\mathbf{2}, \mathbf{2}) \times  (\mathbf{2}, \mathbf{2}) \right]_{sym}  =\left[ (\mathbf{1}, \mathbf{1}) \oplus (\mathbf{3}, \mathbf{3}) \right]_{sym} 
.\eeq
Thus the only way one can construct a scalar from the symmetric part of the  tensor product between two vectors and two self-dual two-forms is from the $(\mathbf{1}, \mathbf{1})$ in both cases above. The factor $\frac{1}{4}$ can be obtained by a back-of-an-envelope calculation.

Another relation that will be useful for later is
  \begin{align}
  \left( -  \tilde{\chi}^{\kappa \lambda} [B_{j k},  \chi^{\rho \sigma} ] 
+  \chi^{\kappa \lambda} [B_{j k}, \tilde{\chi}^{\rho \sigma} ]\right)
        \bar{e}^{+} \Gf \G_\kappa  \G_\lambda  \G_\rho \G_\sigma  \G_j\G_k e^{+}  =16 B_{\mu \nu} [\tilde{\chi}^{\mu \gamma}, \chi^{\nu \gamma}]
. \end{align}
This can be derived by repeated use of  equation \ref{gammas_between_e} together with the self-duality property of the two-forms.
It will be used both in obtaining the fermionic part of the action as well as when computing the supersymmetry variations for the fermionic two-forms.

We can also relate terms containing four copies of the fermionic fields $\phi_i$ to a term quadratic in the cross-product of two $B$'s, such as
\begin{align}
\label{phis_in_B_cross}
[\phi_i, \phi_j] [\phi^i, \phi^j] 
 = &
 \frac{1}{8} (B \times B)_{\mu \nu} (B \times B)^{\mu \nu}
.\end{align}
This can be found by explicitly writing down the expression for the commutator in the adjoint representation of the gauge group and use the properties of the structure constants. This expression will be useful when computing the bosonic terms in the action.

\section{The action in flat space}

The above considerations will allow us to write down the explicit expression for the action of this topologically twisted theory. When performing the twist, we started by considering a dimensional reduction from ten to five dimensions, after which we created our group homomorphism and thus the twist itself. We also found expressions for how to reinterpret the fields from the ten-dimensional theory in terms of our new fields in the twisted theory. It is thus quite logical that we when obtaining the action for this twisted theory again start from the ten-dimensional expression of the action of super-Yang-Mills theory in Minkowski signature, that is from equation \ref{ten_dim_action}.

The Lagrangian density for the topologically twisted theory in five dimensions can then be written as
\beq
\label{action}
\mathcal{S}=\int _{M_4 \times I} \text{d}^4x \text{d}y \text{Tr } \left( \mathcal{L}^{\text{Bosonic}}+\mathcal{L}^{\text{Fermionic}} \right)
.\eeq
In the following sections, we will perform the necessary calculations needed to find both the bosonic piece of this and the fermionic piece. In the next chapter we shall then investigate the supersymmetry invariance of the action obtained herein.

\subsection{The bosonic part}
We will begin by considering the terms containing the bosonic degrees of freedom in the action.  These can by quite straight-forward calculations and use of the expression in equation \ref{phis_in_B_cross} be written as:
\begin{align}
\mathcal{L}^{\text{Bosonic}}=&-\frac{1}{4}F_{\mu \nu} F^{\mu \nu}-\frac{1}{2}F_{y \mu} F^{y \mu}
\\ \nonumber
& -  \frac{1}{8} D_\gamma B_{\mu \nu} D^\gamma B^{\mu \nu}   -   D_\mu \sigma D^\mu\bar{\sigma}  
\\ \nonumber
& -   \frac{1}{8} D_y B_{\mu \nu} D^y B^{\mu \nu}  -  D_y \sigma D^y \bar{\sigma} 
\\ \nonumber
& - \frac{1}{32}(B\times B)_{\mu \nu}(B\times B)^{\mu \nu} - \frac{1}{4} [B_{\mu \nu}, \sigma][B_{\mu \nu}, \bar {\sigma}]  + \frac{1}{2} [\sigma,\bar{\sigma}]^2
.\end{align}
Some of these terms were previously presented in \cite{Witten2011}.

\subsection{The fermionic part}
The fermionic terms are somewhat trickier to obtain.
Since we have managed to write down the ten-dimensional fermionic degrees of freedom in terms of our twisted fields and our constant base element $e^+$, we can now use this to rewrite the fermionic contribution to equation \ref{ten_dim_action}.
By using the expressions in equation \ref{lambda} and \ref{lambda_bar}, together with equation \ref{basic_prop_gammas} and \ref{su2_inv_e}, one can now compute the fermionic part of the Lagrangian density. This will be slightly tedious, but overall straight forward. Let $\mu, \nu$ denote indices on $ M_4$ and as previously $y$ denote the linear coordinate along the interval. After some work, one will find:

\begin{align}
\mathcal{L}^{\text{Fermionic}}
    =& 
\frac{i}{2} \left[ 
  \tilde{\eta}D_\mu \psi^\mu 
    +\eta D_\mu \tilde{\psi}^\mu 
+ \tilde{\psi}^\mu D_\mu \eta 
 +\psi^\mu D_\mu  \tilde{\eta} 
  \right. \\ \nonumber &
+ \tilde{\psi}_\mu D_\nu   \chi^{\mu \nu} 
+  \psi_\mu D_\nu \tilde{\chi}^{\mu \nu}   
 +   \tilde{\chi}^{\mu \nu}D_\nu \psi_\mu 
 +  \chi^{\mu \nu}  D_\nu \tilde{\psi}_\mu 
  \\ \nonumber & 
-\tilde{\eta} D_y \eta - \eta D_y \tilde{\eta} -\tilde{\psi}_\mu D_y \psi^\mu - \psi_\mu D_y \tilde{\psi}^\mu
 -\frac{1}{4}   \tilde{\chi}_{\mu \nu} D_y \chi^{\mu \nu}  -  \frac{1}{4}  \chi_{\mu\nu}D_y\tilde{\chi}^{\mu \nu} 
\\ \nonumber & \left.
\sqrt{2} \left( 
 +  \sigma [ \tilde{\eta} , \tilde{\eta} ]   
 +   \bar{\sigma} [\eta,    \eta ]
 -  \sigma [ \tilde{\psi}_\mu,  \tilde{\psi}^\mu ]   
 -  \bar{\sigma} [\psi_\mu, \psi^\mu ]
 +\frac{1}{4}    \sigma [ \tilde{\chi}^{\mu \nu} ,  \tilde{\chi}_{\mu \nu}  ]
 + \frac{1}{4}    \bar{\sigma} [ \chi^{\mu \nu},  \chi_{\mu \nu} ] \right)
 \right]
\\ \nonumber &
-\frac{1}{4}  B_{\mu \nu} [ \tilde{\eta},   \chi^{ \mu \nu}] 
+\frac{1}{4}  B_{ \mu \nu} [\eta , \tilde{\chi}^{\mu \nu}]  
+   B_{\mu \nu} [ \psi^\mu  , \tilde{\psi}^\nu ]
+ \frac{1}{4} B_{\mu \nu} [\tilde{\chi}^{\mu \gamma}, \chi^{\nu \gamma}]
.\end{align}

In bulk, one can use integration by parts, which can be used to simplify the fermionic part of the Lagrangian density slightly. It should be noted here that this is only valid in bulk, since we otherwise would find surface terms which would not all be identically zero. We will however not concern ourselves with this at the moment, but stay in the bulk where the fermionic part of the Lagrangian density now can be written as
\begin{align}
\mathcal{L}^{\text{Fermionic}}
    =& 
i \left[ 
  \tilde{\eta}D_\mu \psi^\mu 
    +\eta D_\mu \tilde{\psi}^\mu 
+ \tilde{\psi}_\mu D_\nu   \chi^{\mu \nu} 
+  \psi_\mu D_\nu \tilde{\chi}^{\mu \nu}   
-\tilde{\eta} D_y \eta -\tilde{\psi}_\mu D_y \psi^\mu 
 -\frac{1}{4}   \tilde{\chi}_{\mu \nu} D_y \chi^{\mu \nu}   
 \right.
\\ \nonumber & \left.
\frac{1}{\sqrt{2}} \left( 
 +  \sigma [ \tilde{\eta} , \tilde{\eta} ]   
 +   \bar{\sigma} [\eta,    \eta ]
 -  \sigma [ \tilde{\psi}_\mu,  \tilde{\psi}^\mu ]   
 -  \bar{\sigma} [\psi_\mu, \psi^\mu ]
 +\frac{1}{4}    \sigma [ \tilde{\chi}^{\mu \nu} ,  \tilde{\chi}_{\mu \nu}  ]
 + \frac{1}{4}    \bar{\sigma} [ \chi^{\mu \nu},  \chi_{\mu \nu} ] \right)
 \right]
\\ \nonumber &
-\frac{1}{4}  B_{\mu \nu} [ \tilde{\eta},   \chi^{ \mu \nu}] 
+\frac{1}{4}  B_{ \mu \nu} [\eta , \tilde{\chi}^{\mu \nu}]  
+   B_{\mu \nu} [ \psi^\mu  , \tilde{\psi}^\nu ]
+ \frac{1}{4} B_{\mu \nu} [\tilde{\chi}^{\mu \gamma}, \chi^{\nu \gamma}]
.\end{align}
It should be noted that, as opposed to the result for the bosonic part, these terms have never previously been presented.

\section{Supersymmetry}  
In order to show supersymmetry invariance of the action in equation \ref{action}, we must first compute how the fields in the action transform under our supersymmetries. This will, as done when computing the action, be done by starting from the ten-dimensional expressions for the supersymmetry variations of the fields (equations \ref{ferm_variations} and \ref{10_dim_bos}). Let us first start by considering the supersymmetry variations of the bosonic quantities.

\subsection{Variation of the Bosonic fields}
In order to find these,  we need to find the expression for the Dirac conjugate of our supersymmetry parameter. Recall that this  was given by equation \ref{susys}, and thus allows us  to write down the expression for its Dirac conjugate as  
\begin{align}
\bar{\varepsilon}= & \bar{e}^+ (v-u \frac{1}{\sqrt{2}}\Gf \Gamma_{8+i9}) 
.\end{align}
By using \ref{lambda}, we now divide the ten-dimensional expression (equation \ref{10_dim_bos}) into equations relating zero-forms, one-forms and self-dual two-forms respectively, and furthermore split this into smaller pieces according to $U(1)$-charge.
After some calculations, this will give us the supersymmetry transformation rules of the bosonic fields as follows:
\begin{align}
\label{bosonic_transf}
\delta A_y =&
  i v \eta  +i u  \tilde{\eta} 
  \\ \nonumber
\delta \sigma =& - i u \sqrt{2} \eta
 \\ \nonumber
\delta \bar{\sigma} =&  - i v \sqrt{2} \tilde{\eta}
\\ \nonumber
\delta A_\mu=& 
iv \psi_\mu + i u \tilde{\psi}_\mu
\\ \nonumber
\delta B^{\mu \nu}
=&   v \chi^{\mu \nu} - u \tilde{\chi}^{\mu \nu} 
.\end{align}

\subsection{Variation of the Fermionic fields}
In order to compute the variations of the fermionic fields, the ten-dimensional expression in equation \ref{ferm_variations} will be split into  six parts: Two equations with  $U(1)$ charge $\pm \frac{1}{2}$ respectively that will give us the variation of the fermionic zero-forms, and similarly two equations describing the variations of the fermionic one-forms as well as two equations for the variation of the fermionic self-dual two-forms. This will be slightly less straight forward than in the bosonic case.

First, let us start by considering the right hand side of equation \ref{ferm_variations} and divide this in accordance with the above statements. This will be done by noting
\begin{align}
\frac{1}{2} F_{IJ} \G^{IJ}=& 
\frac{1}{2}F_{\mu \nu} \G^\mu\G^\nu+\frac{1}{2} F_{ij} \G^{i+4}\G^{j+4}
 +F_{y i} \G^4 \G^{i+4}
+F_{i\alpha} \G^{i+4} \G^\alpha 
\\ \nonumber &
F_{y \mu} \G^4 \G^\mu+ F_{i \mu} \G^{i+4} \G^\mu +F_{\alpha \mu} \G^\alpha \G^\mu
\\ \nonumber &
+F_{8 9} \G^8 \G^9
+F_{y \alpha} \G^4 \G^\alpha
.\end{align}
The first row in the above expression will give us the right hand side in the equation describing the supersymmetry variation of the  (self-dual) two-forms ($\chi_{\mu \nu}, \tilde{\chi}_{\mu \nu}$) , the second row of the  one-forms, ($\psi_\mu \tilde{\psi}_\mu$), and finally the third row the variation of  the zero-forms ($\eta, \tilde{\eta}$).

The variations under the supersymmetries for the fermionic quantities in the twisted theory can then be calculated in a straight forward manner by using equations \ref{susys} and \ref{lambda} together with \ref{basic_prop_gammas} .  One will then arrive at an expression for the  variations of the fermionic fields under the two scalar supersymmetries as:

\begin{align}
\label{fermionic_SUSYs}
\delta \eta =&  u  [\sigma, \bar{\sigma}]  
-  v\sqrt{2} D_y \sigma
\\ \nonumber
\delta \tilde{\eta} =&
- v   [\sigma, \bar{\sigma}] 
 - u \sqrt{2} D_y \bar{ \sigma}
 \\ \nonumber
 \delta \psi_\nu 
=&
u  F_{y \nu}  - v \sqrt{2}  D_\nu \sigma   
+ i u  D_\mu B_{\nu \mu} 
\\ \nonumber
 \delta \tilde{\psi}_\nu  =&
 v  F_{y \nu}   -u\sqrt{2} D_\nu \bar{\sigma} 
- i v D_\mu B_{\nu \mu}  
\\ \nonumber
\delta \chi_{\kappa \lambda} 
 = &
-2 u F^+_{\kappa \lambda} + u \frac{1}{2} (B \times B)_{\kappa \lambda} +   i \left(u D_y B_{\kappa \lambda}  - v\sqrt{2} [B_{\kappa \lambda}, \sigma] \right)
\\ \nonumber
 \delta \tilde{\chi}_{\kappa \lambda} 
 = &
  -2v F^+_{\kappa \lambda} + v \frac{1}{2}(B \times B)_{\kappa \lambda}   -   i \left(v D_y B_{\kappa \lambda}   - u\sqrt{2} [B_{\kappa \lambda} ,  \bar{\sigma}] \right)
.\end{align}
Again, recall that $F^+_{\kappa \lambda}$ denotes the self-dual part of the field strength for the gauge fields. That only the self-dual part should arise here is obvious since $\chi_{\kappa \lambda}, \tilde{\chi}_{\kappa \lambda}$ are self-dual.

\subsection{Supersymmetry invariance in flat space}
The invariance under the supersymmetries of the obtained action in equation \ref{action} can now be shown. 
It should then be noticed that when performing the variation of this action, one must recall that the variation of a covariant derivative in itself is non-vanishing, so
\beq
\delta (D_\mu \Phi) =  D_\mu \delta \Phi + [\delta A_\mu, \Phi]
.\eeq
It will also be useful to use the below expression for the supersymmetry variation of the field strength for the gauge fields:
\beq
\delta F_{\mu \nu } = D_{[\mu} \delta A_{\nu]} 
.\eeq

Furthermore, on several occasions in the calculation of the variation under the supersymmetries of the action, it will be convenient to use the relation between the commutator of two covariant derivatives an the field strength of the gauge fields, namely:
\beq
\label{flat_comm}
[D_\mu, D_\nu] =F_{\mu \nu}
,\eeq
which is valid when $M_4$ is flat.  By quite extensive calculations, and repeated use of equation \ref{twoformprop}, \ref{flat_comm} and similar expressions together with integrations by parts as well as the Bianchi- and  Jacobi identities, one can eventually show that the obtained action in equation \ref{action} is invariant under the supersymmetries in bulk for $M_4$ with vanishing curvature.

\section{The action in curved space}

If the manifold $M_4$ instead is curved, the above calculation will not hold and some modifications to both the expression for the action (\ref{action}) and the expressions for the variations of the fields (\ref{bosonic_transf},\ref{fermionic_SUSYs}) may be necessary to maintain invariance of the action under the supersymmetries. There are however strong restrictions on which kind of terms that may be added to these because of the requirements posed by for example dimensionality and $U(1)$-charge. We will below see that the variations of the fields will be unchanged from the previous case when $M_4$ was flat, whereas some new terms will be added to the action.

The fault in the calculations in section 4 when $M_4$ is curved comes for the fact that the expression \ref{flat_comm} is no longer true. One must modify this to account for the fact that the covariant derivative no longer will be covariant only with respect to the gauge fields. 
In addition to the term containing the field strength for the gauge fields, each commutator will also give rise to a term proportional to the Riemann tensor. This term will look slightly different depending on what it acts upon. Let $V^\kappa$ be some vector, and consider only the curvature part of the commutator of two covariant derivatives acting on $V^\kappa$. This will be given by:
\begin{align}
\left( [D_\gamma, D_\mu] V^\kappa\right)_{curvature}=& -\frac{1}{2} R_{\gamma \mu \rho \sigma} (\Sigma^{\rho \sigma})^\kappa_{\hspace{2mm} \lambda} V^\lambda
,\end{align}
where $(\Sigma^{\rho \sigma})^a_{\hspace{2mm} b}$ is the generators of the Lorentz group of $M_4$ ($SO(4)$) in the vector representation. These can explicitly be written down such that:
\beq
(\Sigma^{\rho \sigma})^\kappa_{\hspace{2mm} \lambda} = -\delta^{\rho \kappa}\delta^{\sigma}_\lambda + \delta^\rho_\lambda \delta^{\sigma \kappa}
.\eeq

If we instead have a two-form $\Omega^{\kappa \lambda}$ upon which the commutator acts, we will obtain one term for each of the indices $\kappa, \lambda$, so one finds:
\beq
\left( [D_\gamma, D_\mu] \Omega^{\kappa \lambda}\right)_{curvature}= -\frac{1}{2} R_{\gamma \mu \rho \sigma} ( (\Sigma^{\rho \sigma})^\kappa_{\hspace{2mm} \nu} \Omega^{\nu \lambda} + (\Sigma^{\rho \sigma})^\lambda_{\hspace{2mm} \nu} \Omega^{\kappa \nu})
.\eeq

 By taking this into account when performing the variation of the action obtained in \ref{action}, (with the variations of the fields as previously given by equations \ref{bosonic_transf} and  \ref{fermionic_SUSYs}), and again using equation \ref{twoformprop} and similar expressions together with the Bianchi identity, Jacobi identity and known properties of the Riemann tensor, one can eventually show that 
\begin{align}
\label{curved_var}
\delta \mathcal{S}
=&
\int _{M_4 \times I} \text{d}^4x \text{d}y \text{Tr } \left( -\frac{1}{4} R_{\mu \nu \rho \sigma} B^{\mu \nu}  ( v   \chi^{\rho \sigma} -u  \tilde{\chi}^{\rho \sigma}   )
+\frac{1}{8}R B^{\mu \nu} (v   \chi_{\mu \nu}- u  \tilde{\chi}_{\mu \nu} ) \right)
,\end{align}
in bulk. Thus the obtained action in \ref{action} together with the expressions in \ref{bosonic_transf} and \ref{fermionic_SUSYs} cannot be the whole story whenever $M_4$ has non-vanishing curvature. As we mentioned at the beginning of this section though, when $M_4$ is curved the option exists of adding more terms in these equations, and below we shall determine what these correction terms will be. Let us first start by considering corrections to the action presented in equation \ref{action}.

When one consider dimensionality, $U(1)$-charge and the requirement of general covariance, one find that there are only three possible terms that may be added to the action in equation \ref{action} when $M_4$ has non-vanishing curvature. These are:
 \begin{align}
 &
 R \bar{\sigma}\sigma \\ \nonumber &
 RB_{\mu \nu}B^{\mu \nu} \\ \nonumber&
 R_{\mu \nu \rho \sigma} B^{\mu \rho} B^{\nu \sigma}
. \end{align}
 These terms will be added to the original action (equation \ref{action}) with some prefactors which are determined by requiring vanishing of the right hand side in equation \ref{curved_var}. We thus wish to find expressions for how these terms behave under the supersymmetry variation. 
One can here in a straight-forward manner use  the expressions in \ref{bosonic_transf} to find that
  \begin{align}
\delta(  R \bar{\sigma}\sigma) =& - i \sqrt{2} R (u \eta \bar{\sigma}- v \tilde{\eta} \sigma) \\ \nonumber 
 \delta(   RB_{\mu \nu}B^{\mu \nu} ) =& 2 R B_{\mu \nu}(v \chi^{\mu \nu}-u \tilde{\chi}^{\mu \nu})
  \\ \nonumber
 \delta(   R_{\mu \nu \rho \sigma} B^{\mu \rho} B^{\nu \sigma} )=& 2 R_{\mu \nu \rho \sigma} B^{\mu \rho}(v \chi^{\nu \sigma}-u \tilde{\chi}^{\nu \sigma})
. \end{align}
 
By then requiring the expression in \ref{curved_var} to vanish, it is clear that one must add the following terms to the action in equation \ref{action}:  
  \begin{align}
 &
 -\frac{1}{16}RB_{\mu \nu}B^{\mu \nu} \\ \nonumber&
 +\frac{1}{8}R_{\mu \nu \rho \sigma} B^{\mu \rho} B^{\nu \sigma}
. \end{align}
 The first one of the possible term mentioned,  $R \bar{\sigma}\sigma$, cannot exist in the action since its variation would not be cancelled by anything.

Furthermore, the only corrections possible in the expressions for the variations of the fields when $M_4$ has non-vanishing curvature is in the expressions for the variations of the fermionic zero-forms. The only possibility is that there may be an extra term proportional to the curvature scalar added in these (\ref{fermionic_SUSYs}). This will not be the case here however, since it would result in non-vanishing terms in the variation of the action. For example, it will give rise to two terms containing one covariant derivative, one fermionic one-form and the curvature scalar. These two terms will be:
 \beq
 vRD_\mu \psi^\mu+uRD_\mu \tilde{\psi}^\mu
 .\eeq 
 These are not identically zero and are as previously mentioned the only terms of this form.  
 
 Thus with all these considerations in mind, we can draw the conclusion that we have found the complete and unique action of the topologically twisted maximally supersymmetric Yang-Mills theory on a general five manifold on the form $M_4 \times I$ in bulk. This will be given by:

 \begin{align}
 \label{fullAction}
\mathcal{S}
=&
\int _{M_4 \times I} \text{d}^4x \text{d}y \text{Tr } \left(
-\frac{1}{4}F_{\mu \nu} F^{\mu \nu}-\frac{1}{2}F_{y \mu} F^{y \mu}
\right.
\\ \nonumber
& -  \frac{1}{8} D_\gamma B_{\mu \nu} D^\gamma B^{\mu \nu}   -   D_\mu \sigma D^\mu\bar{\sigma}  
\\ \nonumber
& -   \frac{1}{8} D_y B_{\mu \nu} D^y B^{\mu \nu}  -  D_y \sigma D^y \bar{\sigma} 
\\ \nonumber
& - \frac{1}{32}(B\times B)_{\mu \nu}(B\times B)^{\mu \nu} - \frac{1}{4} [B_{\mu \nu}, \sigma][B_{\mu \nu}, \bar {\sigma}]  + \frac{1}{2} [\sigma,\bar{\sigma}]^2
\\ \nonumber
 & 
  -\frac{1}{16}RB_{\mu \nu}B^{\mu \nu} 
 +\frac{1}{8}R_{\mu \nu \rho \sigma} B^{\mu \rho} B^{\nu \sigma}
 \\ \nonumber
 &
i \left[ 
  \tilde{\eta}D_\mu \psi^\mu 
    +\eta D_\mu \tilde{\psi}^\mu 
+ \tilde{\psi}_\mu D_\nu   \chi^{\mu \nu} 
+  \psi_\mu D_\nu \tilde{\chi}^{\mu \nu}   
-\tilde{\eta} D_y \eta -\tilde{\psi}_\mu D_y \psi^\mu 
 -\frac{1}{4}   \tilde{\chi}_{\mu \nu} D_y \chi^{\mu \nu}   
 \right.
\\ \nonumber & \left.
\frac{1}{\sqrt{2}} \left( 
 +  \sigma [ \tilde{\eta} , \tilde{\eta} ]   
 +   \bar{\sigma} [\eta,    \eta ]
 -  \sigma [ \tilde{\psi}_\mu,  \tilde{\psi}^\mu ]   
 -  \bar{\sigma} [\psi_\mu, \psi^\mu ]
 +\frac{1}{4}    \sigma [ \tilde{\chi}^{\mu \nu} ,  \tilde{\chi}_{\mu \nu}  ]
 + \frac{1}{4}    \bar{\sigma} [ \chi^{\mu \nu},  \chi_{\mu \nu} ] \right)
 \right]
\\ \nonumber &
\left.
-\frac{1}{4}  B_{\mu \nu} [ \tilde{\eta},   \chi^{ \mu \nu}] 
+\frac{1}{4}  B_{ \mu \nu} [\eta , \tilde{\chi}^{\mu \nu}]  
+   B_{\mu \nu} [ \psi^\mu  , \tilde{\psi}^\nu ]
+ \frac{1}{4} B_{\mu \nu} [\tilde{\chi}^{\mu \gamma}, \chi^{\nu \gamma}]
\right)
.\end{align}
 
Many of the bosonic terms obtained in this action have previously been obtained by Witten in \cite{Witten2011} during examination of the equations of motion for the theory, but the expression given therein is here completed by the remaining terms.  Neither the bosonic terms consisting of only covariant derivatives and zero-forms or the terms quartic in the zero-forms have been presented before. The same is true for the fermionic terms. The action of the twisted theory is now complete, and the result in equation \ref{fullAction} is the unique result. All coefficients are precisely determined by the condition of invariance under our two supersymmetries.

When considering the boundary as well, surface terms from the integrations by parts that have been carried out here will cause the above action to no longer be invariant under the supersymmetries we have considered here. The boundary conditions of the theory will turn out to be half-BPS \cite{GaiottoWitten2009}, thus breaking half of the supersymmetries. The obtained action will then only be supersymmetric for $u = \pm v$, which have been further investigated in \cite{Witten2011}. However, the precise appearance of the boundary terms have not been studied in any detail in this work. This may be interesting, and possibly something for future works.

 \vspace{1cm}

I would like to thank M\aa ns Henningson for the many illuminating discussions, without which this work would not have been possible.

\newpage
\bibliographystyle{utphys}
\bibliography{references}

\providecommand{\href}[2]{#2}\begingroup\raggedright\begin{thebibliography}{10}

\bibitem{Witten1988}
E.~Witten, ``{Topological Quantum Field Theory},''
  \href{http://dx.doi.org/10.1007/BF01223371}{{\em Commun.Math.Phys.}
  {\bfseries 117} (1988) 353}.

\bibitem{Yamron}
J.~{Yamron}, ``{Topological Actions From Twisted Supersymmetric Theories},''
  {\em Phys.Lett.} {\bfseries B213} (1988) 325.

\bibitem{VafaWitten}
C.~Vafa and E.~Witten, ``{A Strong coupling test of S duality},''
  \href{http://dx.doi.org/10.1016/0550-3213(94)90097-3}{{\em Nucl.Phys.}
  {\bfseries B431} (1994) 3--77},
\href{http://arxiv.org/abs/hep-th/9408074}{{\ttfamily arXiv:hep-th/9408074
  [hep-th]}}.

\bibitem{KapustinWitten}
A.~Kapustin and E.~Witten, ``{Electric-Magnetic Duality And The Geometric
  Langlands Program},'' {\em Commun.Num.Theor.Phys.} {\bfseries 1} (2007)
  1--236,
\href{http://arxiv.org/abs/hep-th/0604151}{{\ttfamily arXiv:hep-th/0604151
  [hep-th]}}.

\bibitem{Marcus1995}
N.~Marcus, ``{The Other topological twisting of N=4 Yang-Mills},''
  \href{http://dx.doi.org/10.1016/0550-3213(95)00389-A}{{\em Nucl.Phys.}
  {\bfseries B452} (1995) 331--345},
\href{http://arxiv.org/abs/hep-th/9506002}{{\ttfamily arXiv:hep-th/9506002
  [hep-th]}}.

\bibitem{Witten2011}
E.~Witten, ``{Fivebranes and Knots},''
  \href{http://arxiv.org/abs/1101.3216}{{\ttfamily arXiv:1101.3216 [hep-th]}}.

\bibitem{Khovanov}
M.~Khovanov, ``{A Categorification Of The Jones Polynomial},'' {\em Duke. Math.
  J.} {\bfseries 101} (2000) 359--426.

\bibitem{Witten2011-2}
E.~Witten, ``{Khovanov Homology And Gauge Theory},''
\href{http://arxiv.org/abs/1108.3103}{{\ttfamily arXiv:1108.3103 [math.GT]}}.

\bibitem{AndersonHenningson}
L.~Anderson and M.~Henningson, ``{Tunneling Solutions in Topological Field
  Theory on $\mathbb{R} \times S^3 \times I$},''
  \href{http://dx.doi.org/10.1007/JHEP02(2012)063}{{\em JHEP} {\bfseries 1202}
  (2012) 063},
\href{http://arxiv.org/abs/1112.2866}{{\ttfamily arXiv:1112.2866 [hep-th]}}.

\bibitem{Henningson2011}
M.~Henningson, ``{Boundary conditions for geometric-Langlands twisted N=4
  supersymmetric Yang-Mills theory},''
  \href{http://dx.doi.org/10.1103/PhysRevD.86.085003}{{\em Phys.Rev.}
  {\bfseries D86} (2012) 085003},
\href{http://arxiv.org/abs/1106.3845}{{\ttfamily arXiv:1106.3845 [hep-th]}}.

\bibitem{GaiottoWitten2011}
D.~Gaiotto and E.~Witten, ``{Knot Invariants from Four-Dimensional Gauge
  Theory},''
\href{http://arxiv.org/abs/1106.4789}{{\ttfamily arXiv:1106.4789 [hep-th]}}.

\bibitem{Henningson2011-2}
M.~Henningson, ``'t~hooft operators in the boundary,''
  \href{http://dx.doi.org/10.1103/PhysRevD.84.105032}{{\em Phys. Rev. D}
  {\bfseries 84} (Nov, 2011) 105032}.
  \url{http://link.aps.org/doi/10.1103/PhysRevD.84.105032}.

\bibitem{Mikhaylov2012}
V.~Mikhaylov, ``{On the Solutions of Generalized Bogomolny Equations},''
  \href{http://dx.doi.org/10.1007/JHEP05(2012)112}{{\em JHEP} {\bfseries 1205}
  (2012) 112},
\href{http://arxiv.org/abs/1202.4848}{{\ttfamily arXiv:1202.4848 [hep-th]}}.

\bibitem{Kaplan2005}
D.~B. Kaplan and M.~Unsal, ``{A Euclidean lattice construction of
  supersymmetric Yang-Mills theories with sixteen supercharges},''
  \href{http://dx.doi.org/10.1088/1126-6708/2005/09/042}{{\em JHEP} {\bfseries
  0509} (2005) 042},
\href{http://arxiv.org/abs/hep-lat/0503039}{{\ttfamily arXiv:hep-lat/0503039
  [hep-lat]}}.

\bibitem{Unsal2006}
M.~Unsal, ``{Twisted supersymmetric gauge theories and orbifold lattices},''
  \href{http://dx.doi.org/10.1088/1126-6708/2006/10/089}{{\em JHEP} {\bfseries
  0610} (2006) 089},
\href{http://arxiv.org/abs/hep-th/0603046}{{\ttfamily arXiv:hep-th/0603046
  [hep-th]}}.

\bibitem{Brink1976}
L.~{Brink}, J.~H. {Schwarz}, and J.~{Scherk}, ``{Supersymmetric Yang-Mills
  Theories},'' \href{http://dx.doi.org/10.1016/0550-3213(77)90328-5}{{\em
  Nucl.Phys.} {\bfseries B121} (1977) 77}.

\bibitem{GaiottoWitten2009}
D.~{Gaiotto} and E.~{Witten}, ``Supersymmetric boundary conditions in
  $\mathcal{N}=4$ super $\textsc{Y}$ang-$\textsc{M}$ills theory,'' {\em Journal
  of Statistical Physics} {\bfseries 135} (2009) 789--855.

\end{thebibliography}\endgroup

\end{document}